\documentclass{icrc29}
\usepackage{unites2e}
\usepackage{graphicx,amssymb,amsmath,times}
\begin{document}
%Title of paper
\title[Status of VERITAS]{Status and Performance of the First VERITAS
telescope} \author[J. Holder et al.] { J. Holder, R.W. Atkins, H. M. Badran,
G. Blaylock, I.H. Bond, P.J. Boyle, S.M. Bradbury,\newauthor J.H. Buckley,
D.A. Carter-Lewis, O. Celik, Y.C.K. Chow, P. Cogan, W. Cui, M.K. Daniel,\newauthor I. de
la Calle Perez, C. Dowdall, P. Dowkontt, C. Duke, T. Ergin, A.D. Falcone,
D.J. Fegan,\newauthor S.J. Fegan, J.P. Finley, L. Fortson, S. Gammell, K. Gibbs,
G.H. Gillanders, J. Grube, J. Hall,\newauthor D. Hanna, E. Hays, D. Horan, S.B. Hughes,
T.B. Humensky, P. Kaaret, G.E. Kenny,\newauthor M. Kertzmann, D.  Kieda, J. Kildea,
J. Knapp, K. Kosack, H. Krawczynski, F. Krennrich,\newauthor M.J. Lang, S. LeBohec,
E. Linton, J. Lloyd-Evans, G. Maier, H. Manseri, A. Milovanovic,\newauthor P. Moriarty,
R. Mukherjee, T.N. Nagai, P.A. Ogden, M. Olevitch, R.A. Ong, J.S. Perkins,\newauthor
D. Petry, F. Pizlo, M. Pohl, B. Power-Mooney, J. Quinn, M. Quinn, K. Ragan,
P.T. Reynolds,\newauthor P. Rebillot, H.J. Rose, M. Schroedter, G.H. Sembroski,
D. Steele, S.P. Swordy, A. Syson,\newauthor L. Valcarcel, V.V. Vassiliev, R.G. Wagner,
S.P. Wakely, G. Walker, T.C. Weekes, R. J. White,\newauthor D.A. Williams, J. Zweerink\\
} \presenter{Presenter: J. Holder (jh@ast.leeds.ac.uk), \
uki-holder-J-abs1-og27-oral}

\maketitle

\begin{abstract}

 The first of the four atmospheric Cherenkov telescopes of the VERITAS array
 has been in operation at the Mt. Hopkins base camp since January 2005. The
 telescope has met all specifications. We present here a description of the
 technical performance, including calibration details and a
 summary of a preliminary analysis of Crab Nebula observations. The
 construction status of the complete VERITAS array is also discussed.

\end{abstract}
\vspace{-0.3cm}
\section{Introduction}

\begin{figure}[h]
  \begin{center}
    \begin{tabular}{cc}
      \includegraphics*[height=4.5cm]{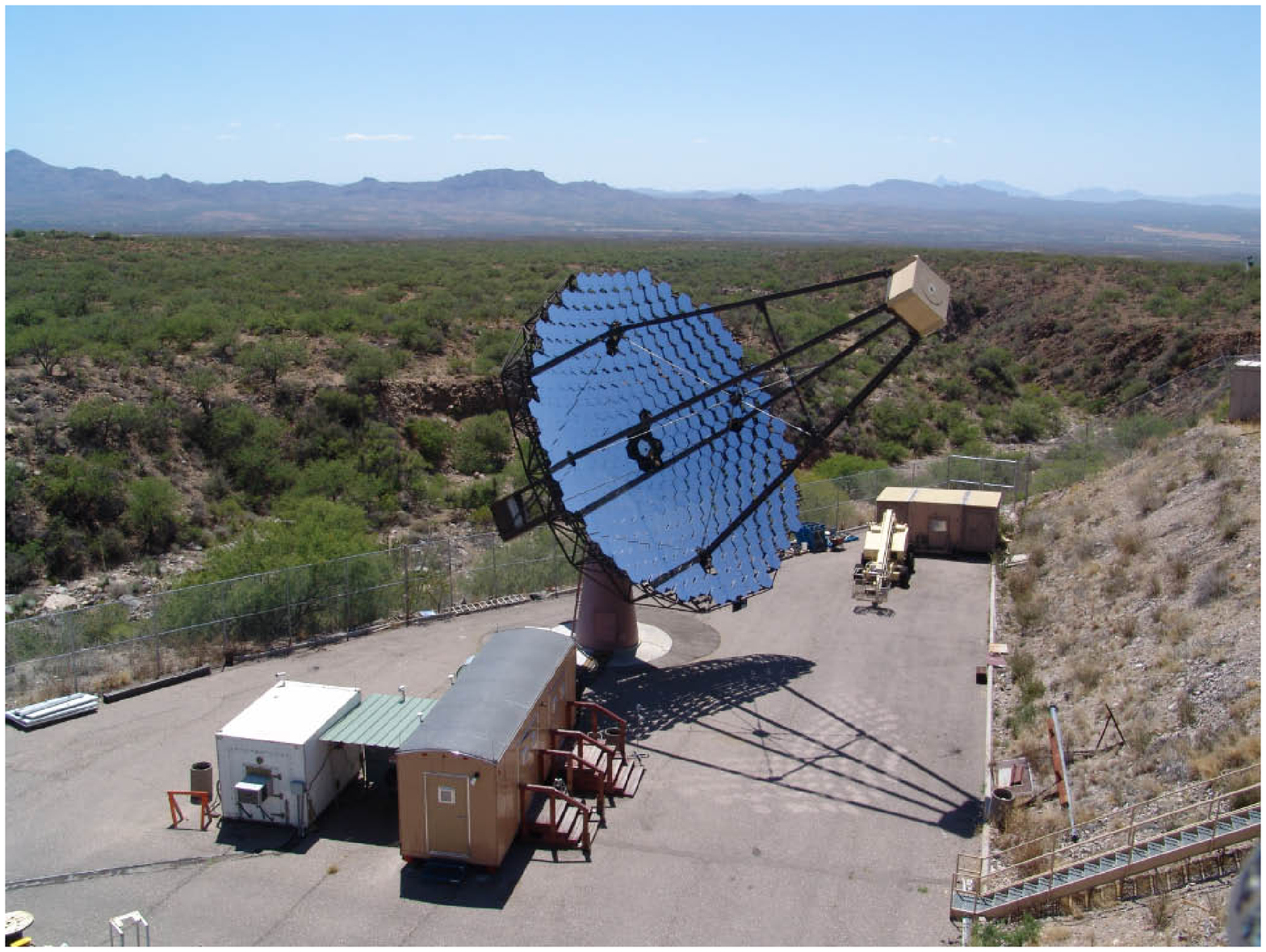}
      \hspace{1cm}
      \includegraphics*[height=4.5cm]{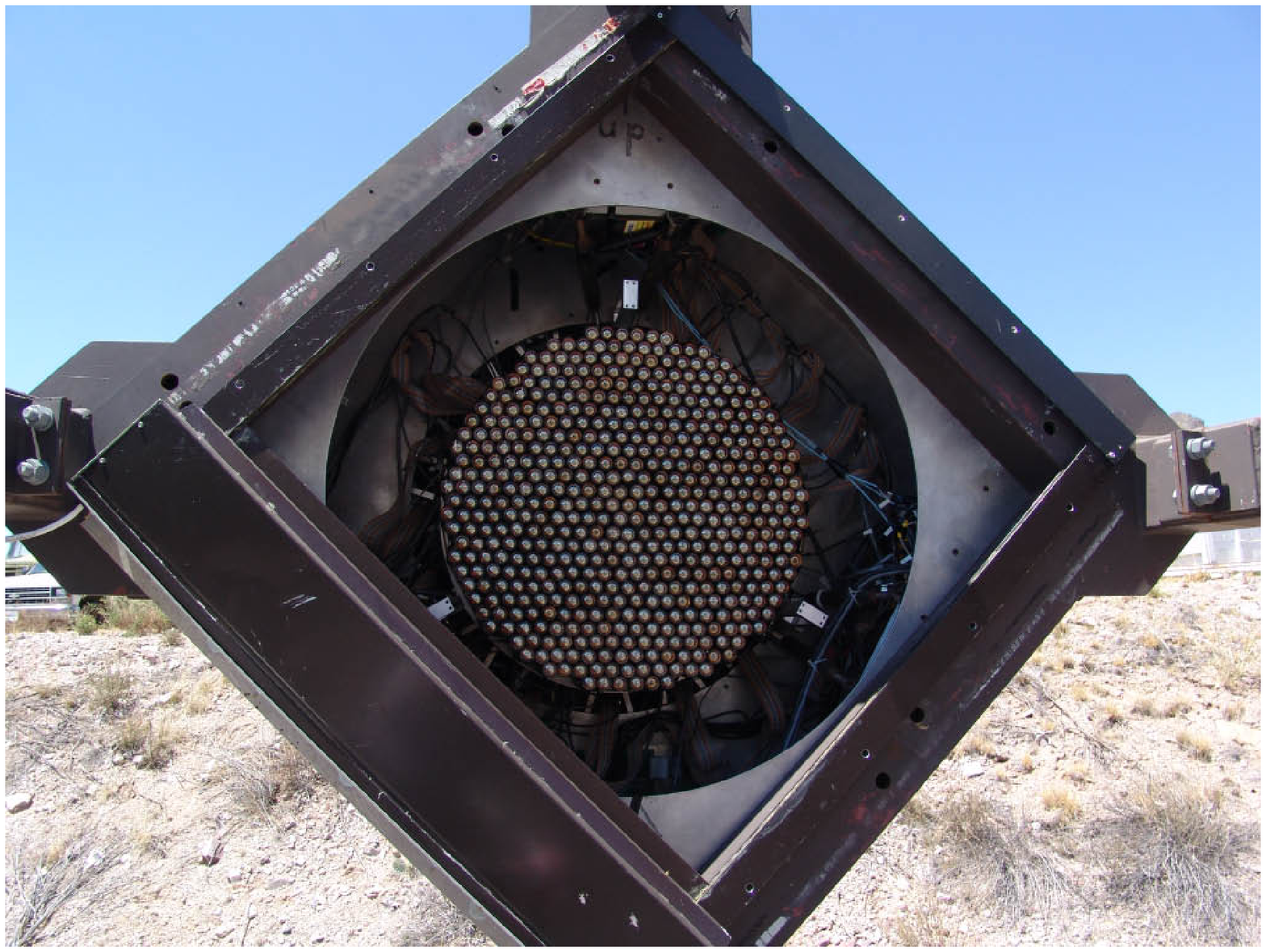}
    \end{tabular}
    \caption{\label{Tel1}
      {\bf Left:} VERITAS Telescope 1. {\bf Right:} The 499 pixel PMT camera.
    }
  \end{center}
\vspace{-0.3cm}
\end{figure}

The VERITAS Collaboration is building an array of imaging atmospheric
Cherenkov telescopes for ground-based gamma-ray astronomy. Phase-I of the
project consists of four, $12\U{m}$ telescopes sited at Horseshoe Canyon on
Kitt Peak, Arizona, at an altitude of $1800\U{m}$. Figure~\ref{Tel1} shows the
first of these telescopes temporarily installed at the base camp of the Whipple
Observatory at Mt. Hopkins (altitude=$1275\U{m}$).

\vspace{-0.3cm}
\section{Mechanical/Optical Performance}

The VERITAS telescopes follow a Davies-Cotton optical design with $12\U{m}$
aperture and $12\U{m}$ focal length. The mechanical structure consists of an
altitude-azimuth mount and a steel optical support structure (OSS). The mount
is a commercial unit manufactured by RPM-PSI (Northridge,
California); the OSS is custom designed by M3 Engineering (Tucson, Arizona)
and fabricated by Amber Steel (Chandler, Arizona) \cite{Gibbs03}.

The tracking is measured to be accurate to $<0.01^{\circ}$ with a maximum
slew speed of $0.3^{\circ}\UU{s}{-1}$. Tests with a slightly modified drive
system have enabled us to reach maximum slew speeds of $1^{\circ}\UU{s}{-1}$;
the remaining telescopes will have this modification installed as standard.

The 350 individual mirror facets are hexagonal, each with an area of
$0.322\UU{m}{2}$, providing a total mirror area of $\sim110\UU{m}{2}$. They
are made from glass, slumped and polished and then coated with aluminium and
anodized at a dedicated facility on-site. Reflectivity is $>90\%$ at
$320\U{nm}$. Each facet is shaped with a $24\U{m}$ radius of curvature and
arranged on a spherical surface of $12\U{m}$ radius. Throughout most of the
observations presented in this paper, the point spread function (PSF) at the
position of Polaris was measured to be $0.09^{\circ}$ FWHM.  Recently, the
telescope alignment has been optimized, and the current PSF is $0.06^{\circ}$.

 A prototype version of the telescope started operations in early 2004,
 and the completed telescope has been operating since January 2005. The
 structure is robust and the positioning precision very reliable. We envisage no problems
 or major changes to the design for the remaining VERITAS telescopes.

\vspace{-0.3cm}
\section{Camera and Electronics}

\begin{figure}[h]
  \begin{center}
    \begin{tabular}{cc}
      \includegraphics*[height=5.cm]{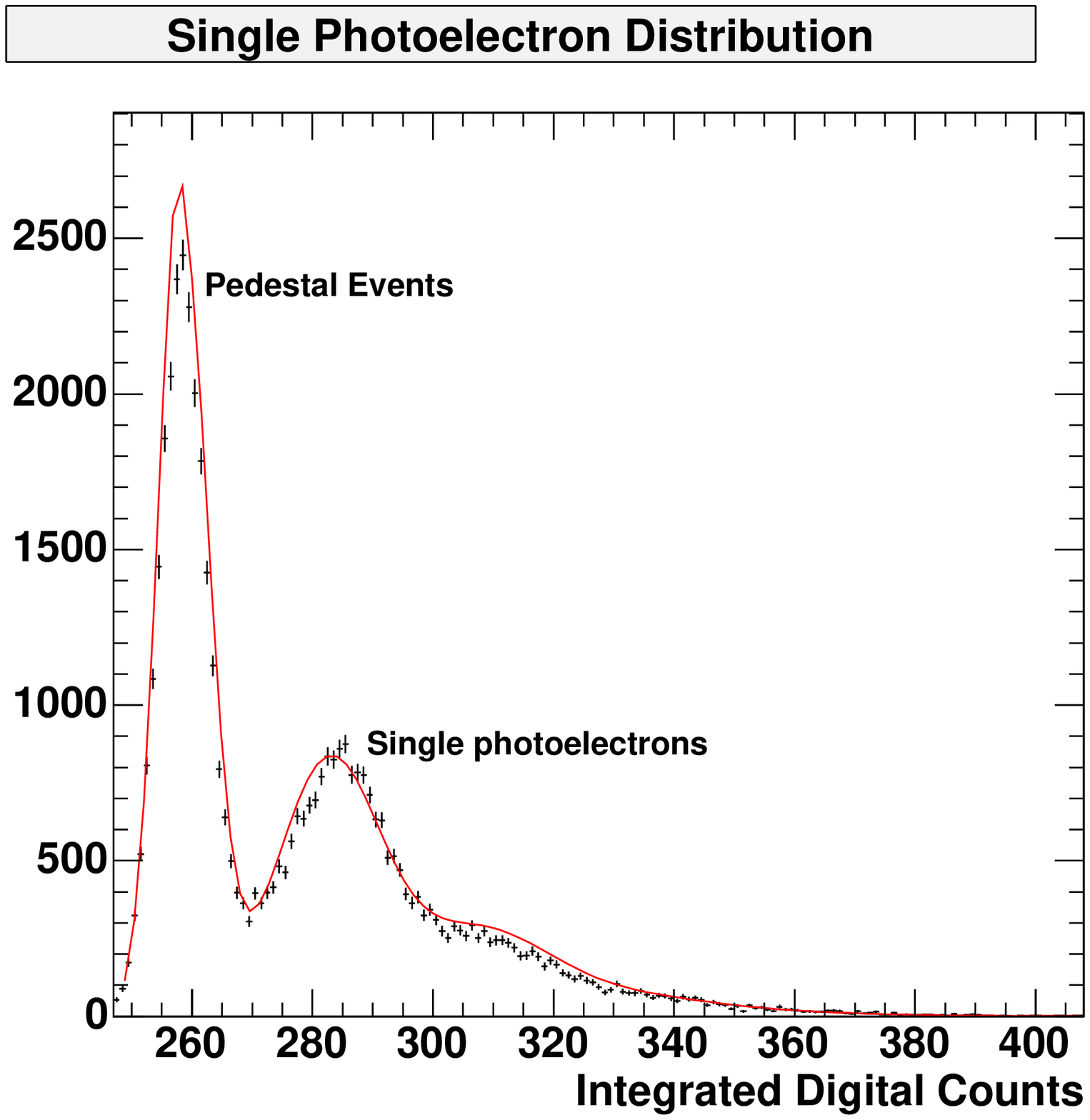}
      \hspace{1cm}
      \includegraphics*[height=5.cm]{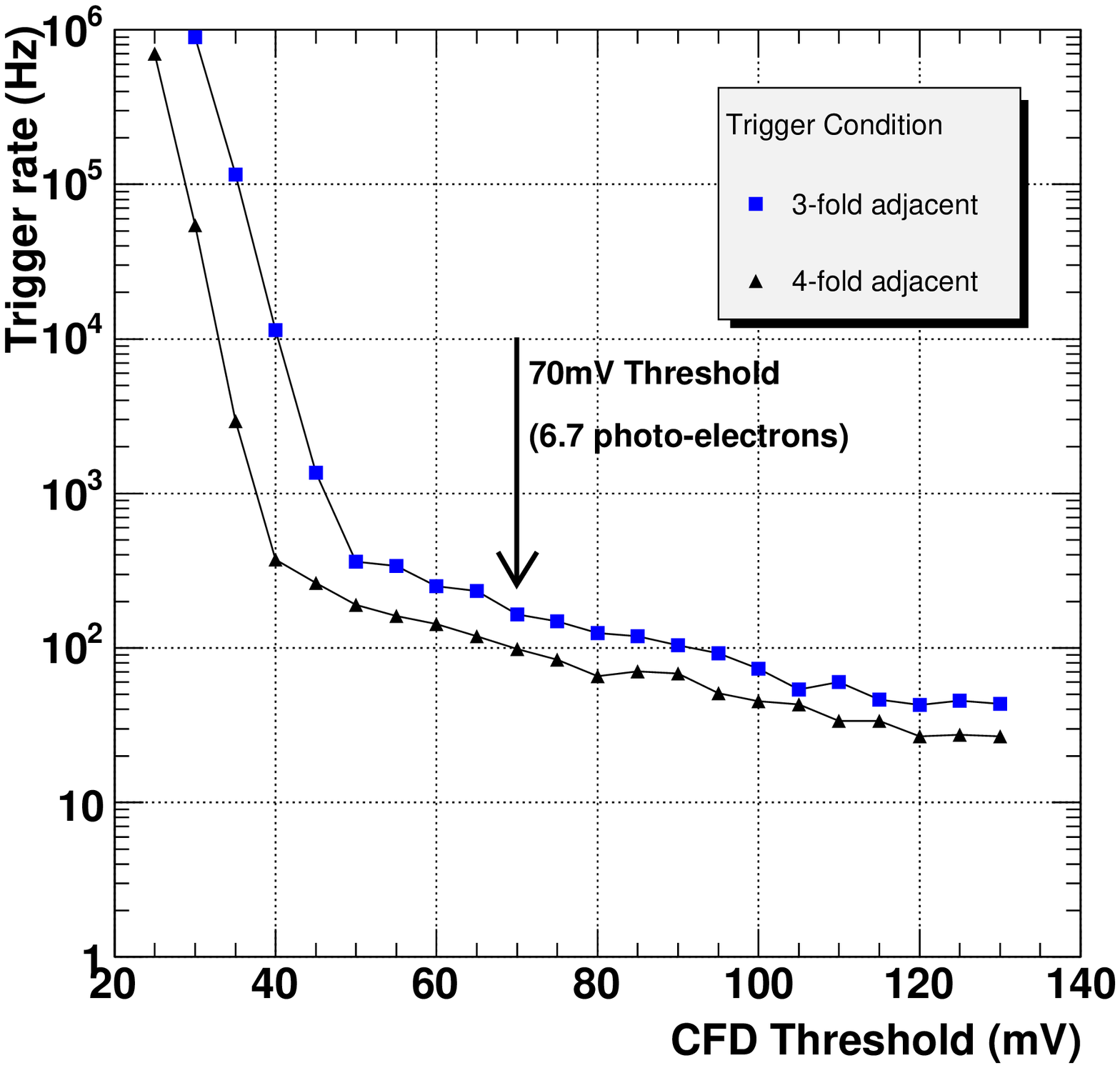}
    \end{tabular}
    \caption{\label {singlepe} {\bf Left:} The single photoelectron response
      for one PMT (at an increased gain of $\sim1\times10^{6}$). The fit
      assumes a Poisson distribution of photoelectrons, and a Gaussian
      distribution for the integrated charge produced by a single p.e. {\bf Right:}
      The trigger rate as a function of CFD threshold for two different
      pattern trigger configurations.  }
  \end{center}
\vspace{-0.3cm}
\end{figure}

The focal plane instrumentation is a 499 element photomultiplier tube (PMT)
camera, with $0.15^{\circ}$ angular spacing giving a field-of-view of
$3.5^{\circ}$. The PMTs are Photonis XP2970/02 with a quantum efficiency
$>20\%$ at $300\U{nm}$, currently running at a gain of
$\sim2\times10^{5}$. Light cones have not yet been installed: two different
designs are being fabricated and one will be installed in Autumn 2005,
significantly increasing our photon collection area. Figure~\ref{singlepe}
shows the single photoelectron response for a single PMT, measured \textit{in
situ} by placing a mylar screen in front of the camera and then illuminating
with a laser pulse.

The PMT signals are amplified by a high-bandwidth preamplifier integrated into
the PMT base mounts. This circuit also allows the PMT anode currents to be
monitored; currents are typically $3~\mu$A (for dark fields) to
$6~\mu$A (for bright fields), corresponding to a night-sky photoelectron
background of 100 - $200\U{MHz}$. The signals are sent via $\sim50\U{m}$
of RG59 stranded cable to the telescope trigger and data acquisition
electronics, at which point the signal pulse for an input delta function has a
risetime (10\% to 90\%) of $3.3\U{ns}$ and a width of $6.5\U{ns}$. The
multi-level trigger system consists of a programmable constant fraction
discriminator (CFD) for each PMT \cite{Hall03}, the output of which is passed
to a pattern recognition trigger system \cite{Bradbury02} which is
programmed to recognise triggers resembling true compact Cherenkov light
flashes. Figure~\ref{singlepe} shows the trigger rate as a function of CFD
threshold for two different pattern trigger configurations. Observations this
year have all been made with a rather conservative threshold of $\sim6-7$
photoelectrons and a 3-fold adjacent pixel pattern trigger configuration,
giving a cosmic ray rate at high elevation of $\sim150\U{Hz}$.

\begin{figure}[h]
  \begin{center}
    \begin{tabular}{cc}
      \includegraphics*[height=5.5cm]{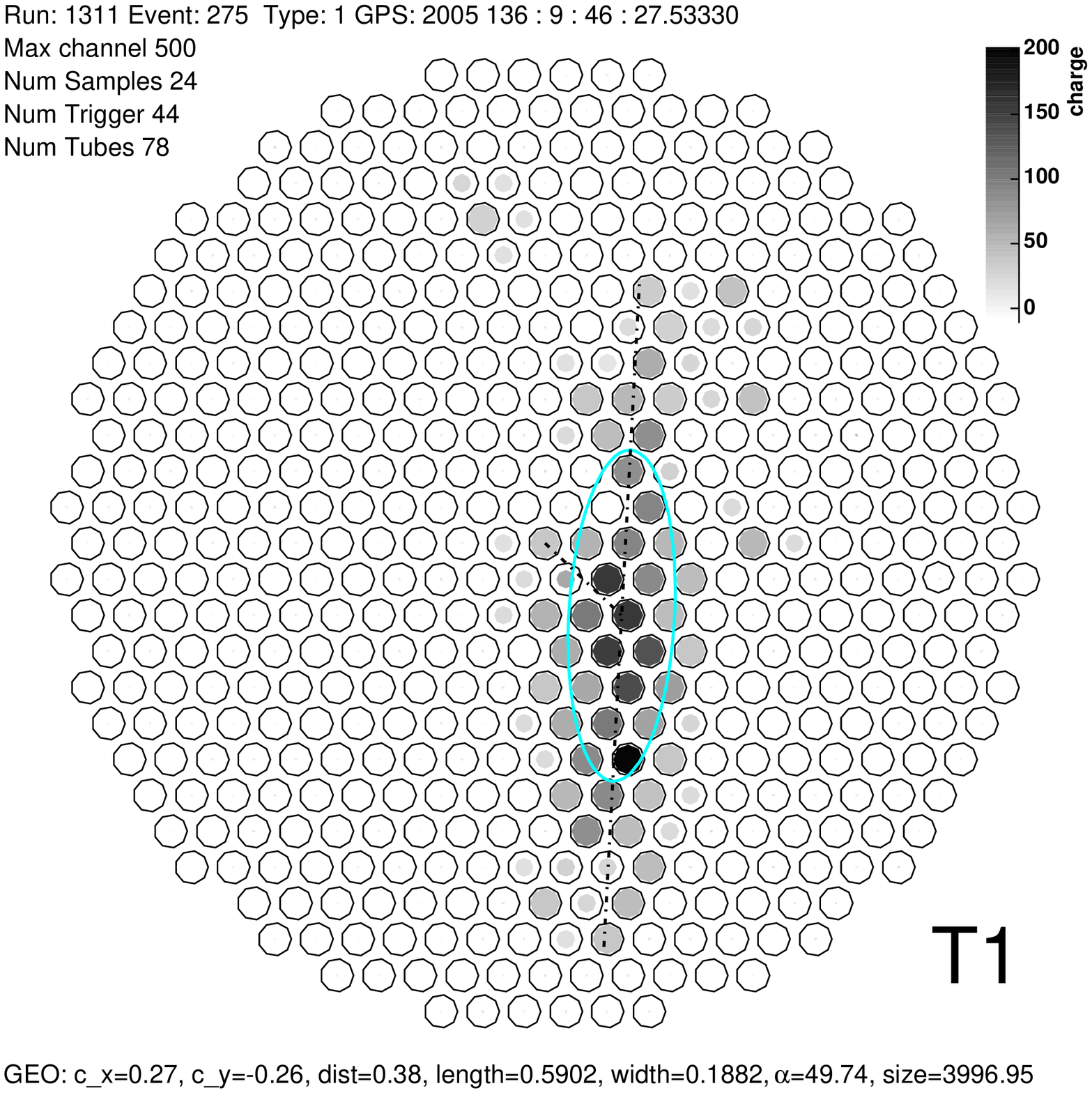}
      \hspace{1cm}
      \includegraphics*[height=5.5cm]{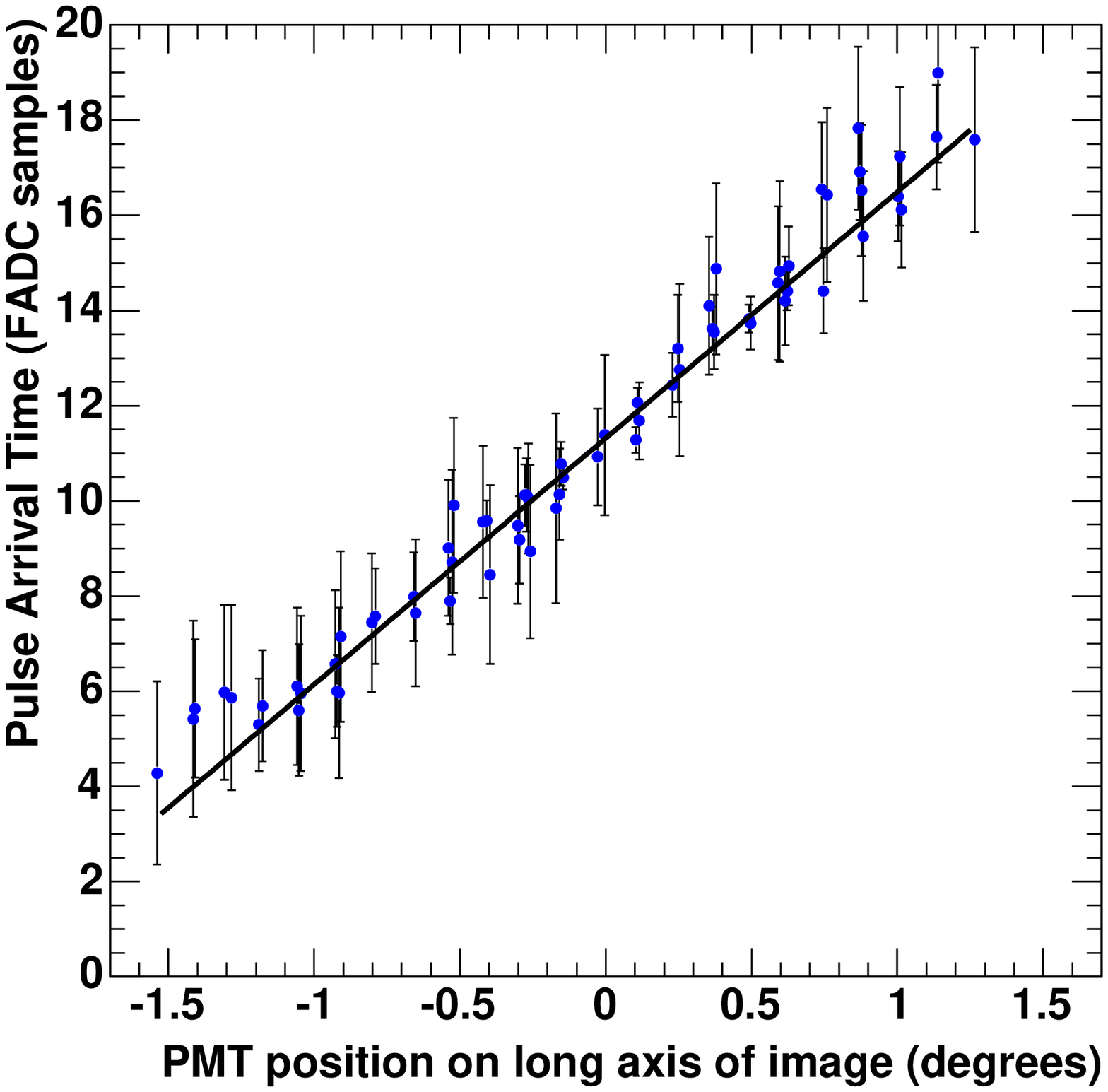}
    \end{tabular}
    \caption{\label {event} {\bf Left:} The charge distribution across the
      camera for a cosmic ray event (the grey scale is in d.c.). {\bf
      Right:} The Cherenkov pulse arrival time distribution (in units of FADC
      samples = $2\U{ns}$) along the long axis of the image on the left.}
  \end{center}
\vspace{-0.3cm}
\end{figure}

The PMT signals are digitized using custom-built VME boards housing Flash ADCs
with $2\U{ns}$ sampling and a memory depth of $32\mu$s \cite{Buckley03}. By
default the signal traces follow a high gain path to the FADC; however, if the
$8\U{bit}$ dynamic range is exceeded, an analog switch connects the FADC chip
to a delayed low gain channel instead, extending the dynamic range for a
single $2\U{ns}$ sample from 256 to 1500 digital counts (d.c.), where
$1\U{d.c}\sim0.19$~photoelectrons at our current PMT gain. The electronic
noise is small, with a sample-to-sample variance of $\sim0.5\U{d.c.}$ and an
event-to-event variance over a 10 sample integration window of
$\sim1.5\U{d.c.}$. The readout window size and position is programmable; a 24
sample window readout on all 500 channels results in a data size of
$13.5\U{kb}$ per event and a deadtime of $\sim10\%$ at $150\U{Hz}$. While this
is manageable for a single telescope, the VERITAS array will produce four
times as much data at higher rates (up to $1\U{kHz}$). To cope with this we
have implemented a zero suppression scheme, whereby only those channels with a
peak charge larger than a preset value are read out, reducing the data size by
a factor of approximately four. As well as allowing us to minimize the charge
integration gate and hence improve the signal/noise, the FADCs also provide
the time distribution of the Cherenkov photons across the image (see
\cite{Holder05}). Figure~\ref{event} shows the charge and arrival time
information for a cosmic ray event with significant time structure.

\vspace{-0.3cm}
\section{Observations and Data Analysis}

\begin{figure}[h]
  \begin{center}
    \begin{tabular}{cc}
      \includegraphics*[height=5.5cm]{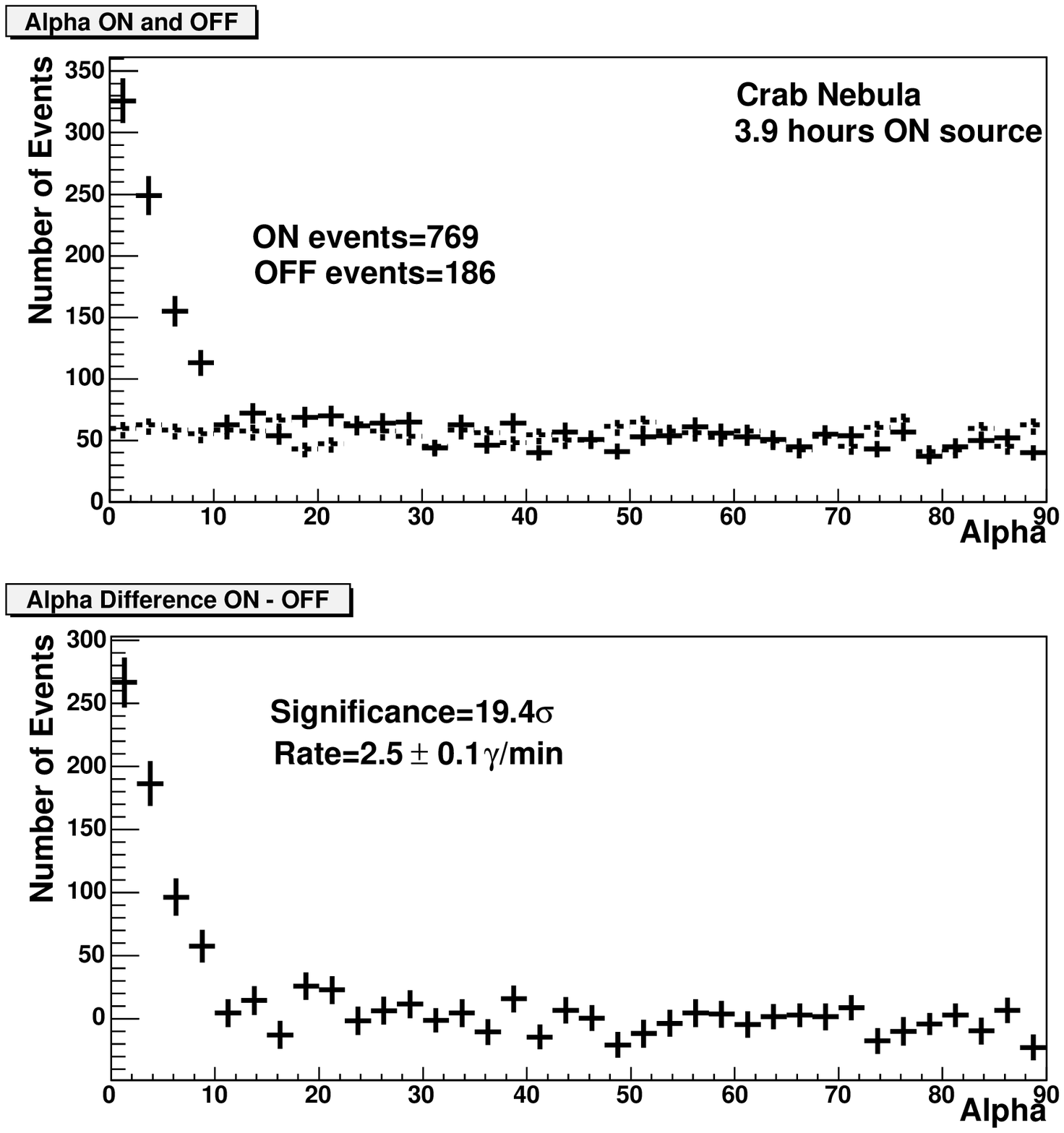}
      \hspace{1cm}
      \includegraphics*[height=5.5cm]{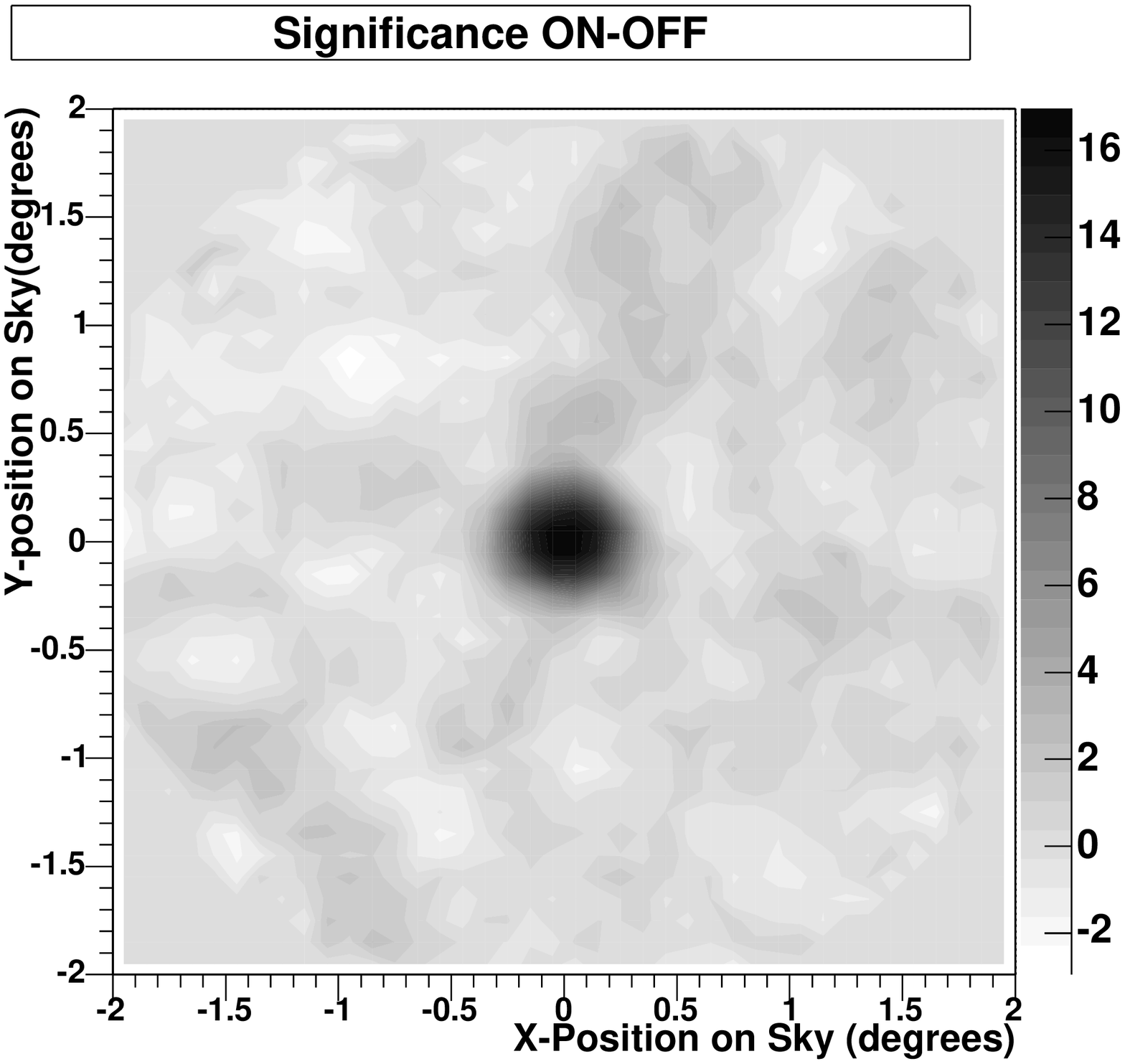}
    \end{tabular}
    \caption{\label {crab}
      {\bf Left:} The image orientation angle, alpha, for ON and OFF source
      observations after gamma-ray selection cuts {\bf Right:}
      The two-dimensional map of reconstructed source position. Adjacent bins are not independent.
    }
  \end{center}
\vspace{-0.3cm}
\end{figure}

Routine observations with the first VERITAS telescope began in February 2005,
in time to collect a small dataset on the Crab Nebula at high
elevation. Observations were taken in the standard ON-OFF mode, and events
parameterized and passed through gamma-ray selection cuts in a similar fashion
to observations with the Whipple 10~m telescope. Figure~\ref{crab} shows the
results for an ON source exposure of $3.9\U{hours}$, indicating a sensitivity
of $\sim10\sigma$ for 1 hour of ON source observations. Also shown is a
significance map of the reconstructed source position using the method of
Lessard et al. \cite{Lessard01} and smoothed with a simple smoothing
algorithm. The rather low gamma-ray rate is the result of hard gamma-ray cuts
which must be applied to reject the local muon background. We have developed a
full simulation chain \cite{Maier05} which indicates a gamma-ray threshold for
the Crab observations of $\sim150\U{GeV}$ (22 $\gamma\UU{min}{-1}$) before
cuts and $\sim370\U{GeV}$ after. Muon rejection will take place at the
hardware trigger level once more than one telescope is installed, removing the
need for hard cuts and dramatically improving the sensitivity. 

An observing campaign, complementary to that of the Whipple 10~m continued
until June 2005, resulting in $>10\sigma$ detections of the known gamma-ray
sources Mkn~421 and Mkn~501 \cite{Cogan05} and data sets on various potential
TeV sources. A full online analysis package enables us to detect strong flaring
behaviour within minutes.

\vspace{-0.5cm}
\section{Summary}
The first VERITAS telescope has been operating throughout 2005, has met all
technical specifications and detected a number of TeV gamma-ray sources. The
Kitt Peak site for VERITAS Phase-I has undergone significant development; site
clearance, power line installation and construction of all four telescope pads
has been completed. The major mechanical components of all four telescopes
have been delivered to the Mt Hopkins base camp. Because of a temporary delay
in access to the Kitt Peak site, we have decided to install the second VERITAS
telescope at the Whipple base camp, $85\U{m}$ away from the first telescope on
an East-West baseline.  By operating these two telescopes together beginning
Autumn 2005, we will be able to reject the muon background and dramatically
increase sensitivity, as well as test the array trigger electronics.

\end{document}